\documentclass[superscriptaddress,preprintnumbers,amsmath,amssymb,twocolumn,floats,aps,prl]{revtex4-1}
\usepackage{txfonts}
\usepackage{amssymb}
\usepackage{graphicx}
\usepackage{epstopdf}
\usepackage{sidecap}
\usepackage{xcolor, appendix}
\usepackage{CJK}
\usepackage{txfonts}
\usepackage{url}

\begin{document}

\bibliographystyle{unsrt}
\title{Dual topological superconducting states in the layered titanium-based oxypnictide superconductor BaTi$_2$Sb$_2$O}

\author{Z. Huang}
\thanks{Equal contributions}
\affiliation{Center for Excellence in Superconducting Electronics, State Key Laboratory of Functional Materials for Informatics, Shanghai Institute of Microsystem and Information Technology, Chinese Academy of Sciences, Shanghai 200050, China}
\affiliation{Center of Materials Science and Optoelectronics Engineering, University of Chinese Academy of Sciences, Beijing 100049, China}
\affiliation{School of Physical Science and Technology, ShanghaiTech University, Shanghai 201210, China}

\author{W. L. Liu}
\thanks{Equal contributions}
\affiliation{Center for Excellence in Superconducting Electronics, State Key Laboratory of Functional Materials for Informatics, Shanghai Institute of Microsystem and Information Technology, Chinese Academy of Sciences, Shanghai 200050, China}
\affiliation{Center of Materials Science and Optoelectronics Engineering, University of Chinese Academy of Sciences, Beijing 100049, China}
\affiliation{School of Physical Science and Technology, ShanghaiTech University, Shanghai 201210, China}

\author{H. Y. Wang}
\thanks{Equal contributions}
\affiliation{School of Physical Science and Technology, ShanghaiTech University, Shanghai 201210, China}

\author{Y. L. Su}
\thanks{Equal contributions}
\affiliation{School of Physical Science and Technology, ShanghaiTech University, Shanghai 201210, China}

\author{Z. T. Liu}
\thanks{Equal contributions}
\affiliation{Center for Excellence in Superconducting Electronics, State Key Laboratory of Functional Materials for Informatics, Shanghai Institute of Microsystem and Information Technology, Chinese Academy of Sciences, Shanghai 200050, China}

\author{X. B. Shi}
\affiliation{State Key Laboratory of Advanced Welding and Joining, Harbin Institute of Technology, Shenzhen 518055, China}
\affiliation{Flexible Printed Electronics Technology Center, Harbin Institute of Technology, Shenzhen 518055, China}

\author{S. Y. Gao}
\affiliation{Beijing National Laboratory for Condensed Matter Physics and Institute of Physics, Chinese Academy of Sciences, Beijing 100190, China}

\author{Z. C. Jiang}
\affiliation{Center for Excellence in Superconducting Electronics, State Key Laboratory of Functional Materials for Informatics, Shanghai Institute of Microsystem and Information Technology, Chinese Academy of Sciences, Shanghai 200050, China}

\author{Z. H. Liu}
\affiliation{Center for Excellence in Superconducting Electronics, State Key Laboratory of Functional Materials for Informatics, Shanghai Institute of Microsystem and Information Technology, Chinese Academy of Sciences, Shanghai 200050, China}
\affiliation{Center of Materials Science and Optoelectronics Engineering, University of Chinese Academy of Sciences, Beijing 100049, China}

\author{J. S. Liu}
\affiliation{Center for Excellence in Superconducting Electronics, State Key Laboratory of Functional Materials for Informatics, Shanghai Institute of Microsystem and Information Technology, Chinese Academy of Sciences, Shanghai 200050, China}

\author{X. L. Lu}
\affiliation{Center for Excellence in Superconducting Electronics, State Key Laboratory of Functional Materials for Informatics, Shanghai Institute of Microsystem and Information Technology, Chinese Academy of Sciences, Shanghai 200050, China}

\author{Y. C. Yang}
\affiliation{Center for Excellence in Superconducting Electronics, State Key Laboratory of Functional Materials for Informatics, Shanghai Institute of Microsystem and Information Technology, Chinese Academy of Sciences, Shanghai 200050, China}

\author{J. X. Zhang}
\affiliation{School of Physical Science and Technology, ShanghaiTech University, Shanghai 201210, China}
\affiliation{Institute for Advanced Study, Tsinghua University, Beijing 100084, China}

\author{S. C. Huan}
\affiliation{School of Physical Science and Technology, ShanghaiTech University, Shanghai 201210, China}

\author{W. Xia}
\affiliation
{School of Physical Science and Technology, ShanghaiTech University, Shanghai 201210, China}
\affiliation{\mbox{ShanghaiTech Laboratory for Topological Physics, ShanghaiTech University, Shanghai 201210, China}}

\author{J. H. Wang}
\affiliation
{School of Physical Science and Technology, ShanghaiTech University, Shanghai 201210, China}
\affiliation{\mbox{ShanghaiTech Laboratory for Topological Physics, ShanghaiTech University, Shanghai 201210, China}}

\author{Y. S. Wu}
\affiliation
{School of Physical Science and Technology, ShanghaiTech University, Shanghai 201210, China}

\author{X. Wang}
\affiliation{School of Physical Science and Technology, ShanghaiTech University, Shanghai 201210, China}
\affiliation{Analytical Instrumentation Center, School of Physical Science and Technology, ShanghaiTech University, Shanghai 201210, China}

\author{N. Yu}
\affiliation{School of Physical Science and Technology, ShanghaiTech University, Shanghai 201210, China}
\affiliation{Analytical Instrumentation Center, School of Physical Science and Technology, ShanghaiTech University, Shanghai 201210, China}

\author{Y. B. Huang}
\affiliation{Shanghai Synchrotron Radiation Facility, Shanghai Advanced Research Institute, Chinese Academy of Sciences, 201204 Shanghai, China}

\author{S. Qiao}
\affiliation{Center for Excellence in Superconducting Electronics, State Key Laboratory of Functional Materials for Informatics, Shanghai Institute of Microsystem and Information Technology, Chinese Academy of Sciences, Shanghai 200050, China}

\author{J. Li}
\affiliation
{School of Physical Science and Technology, ShanghaiTech University, Shanghai 201210, China}
\affiliation{\mbox{ShanghaiTech Laboratory for Topological Physics, ShanghaiTech University, Shanghai 201210, China}}

\author{W. W. Zhao}
\affiliation{State Key Laboratory of Advanced Welding and Joining, Harbin Institute of Technology, Shenzhen 518055, China}
\affiliation{Flexible Printed Electronics Technology Center, Harbin Institute of Technology, Shenzhen 518055, China}

\author{Y. F. Guo}
\email{guoyf@shanghaitech.edu.cn}
\affiliation{School of Physical Science and Technology, ShanghaiTech University, Shanghai 201210, China}

\author{G. Li}
\email{ligang@shanghaitech.edu.cn}
\affiliation{School of Physical Science and Technology, ShanghaiTech University, Shanghai 201210, China}
\affiliation{\mbox{ShanghaiTech Laboratory for Topological Physics, ShanghaiTech University, Shanghai 201210, China}}

\author{D. W. Shen}
\email{dwshen@mail.sim.ac.cn}
\affiliation{Center for Excellence in Superconducting Electronics, State Key Laboratory of Functional Materials for Informatics, Shanghai Institute of Microsystem and Information Technology, Chinese Academy of Sciences, Shanghai 200050, China}
\affiliation{Center of Materials Science and Optoelectronics Engineering, University of Chinese Academy of Sciences, Beijing 100049, China}

\begin{abstract}

Topological superconductors have long been predicted to host Majorana zero modes which obey non-Abelian statistics and have potential for realizing non-decoherence topological quantum computation. However, material realization of topological superconductors is still a challenge in condensed matter physics. Utilizing high-resolution angle-resolved photoemission spectroscopy and first-principles calculations, we predict and then unveil the coexistence of topological Dirac semimetal and topological insulator states in the vicinity of Fermi energy (\emph{E$_F$}) in the titanium-based oxypnictide superconductor BaTi$_2$Sb$_2$O. Further spin-resolved measurements confirm its spin-helical surface states around \emph{E$_F$}, which are topologically protected and give an opportunity for realization of Majorana zero modes and Majorana flat bands in one material. Hosting dual topological superconducting states, the intrinsic superconductor BaTi$_2$Sb$_2$O is expected to be a promising platform for further investigation of topological superconductivity.

\end{abstract}

\maketitle

Topological superconductors have attracted tremendous interest for harboring Majorana bound states on their boundaries~\cite{hatsugai1993chern,teo2010topological,tanaka2011symmetry}. The non-Abelian Majorana zero modes in the vortex of topological superconductors are potential for topological quantum computations without decoherence~\cite{read2000paired,kitaev2003fault,nayak2008non}. 
To date, several systems have been predicted to host topological superconductivity (TSC).
Examples include intrinsic odd-parity superconductors~\cite{mackenzie2003superconductivity,fu2010odd,sato2010topological,hsieh2012majorana} and those heterostructures constructed by proximity coupling of topological insulators (TIs) to conventional \emph{s}-wave superconductors~\cite{fu2008superconducting,qi2010topological,sau2010generic}, which inspired enormous experimental efforts to the exploration of Majorana fermions~\cite{kashiwaya2011edge,jang2011observation,hor2010superconductivity,wray2010observation,sasaki2011topological,matano2016spin,yonezawa2017thermodynamic,liu2015superconductivity,tanaka2013two,williams2012unconventional,wang2012coexistence,wang2013fully,cho2013symmetry,oostinga2013josephson,finck2014phase,hart2014induced,mourik2012signatures,das2012zero,lee2014spin,albrecht2016exponential,nadj2014observation}. However, experimentally, the presence of Majorana zero modes in some of these systems is still in hot debate. For example, pairing symmetries of several proposed \emph{p}-wave superconductors, such as Sr$_2$RuO$_4$~\cite{kashiwaya2011edge,jang2011observation} and Cu$_x$Bi$_2$Se$_3$~\cite{hor2010superconductivity,wray2010observation,sasaki2011topological,matano2016spin,yonezawa2017thermodynamic}, are yet inconclusive. 
On the other side, although signs of Majorana bound states have been reported in conventional \emph{s}-wave superconductors with the proximity to TIs, difficulties in fabricating such these heterostructures and the disturbance of interface physics would inevitably prohibit their further studies~\cite{williams2012unconventional,wang2012coexistence,wang2013fully,cho2013symmetry,oostinga2013josephson,finck2014phase,hart2014induced,mourik2012signatures,das2012zero,lee2014spin,albrecht2016exponential,nadj2014observation}.

\begin{figure*}[htbp]
\centering
\includegraphics[width=17cm]{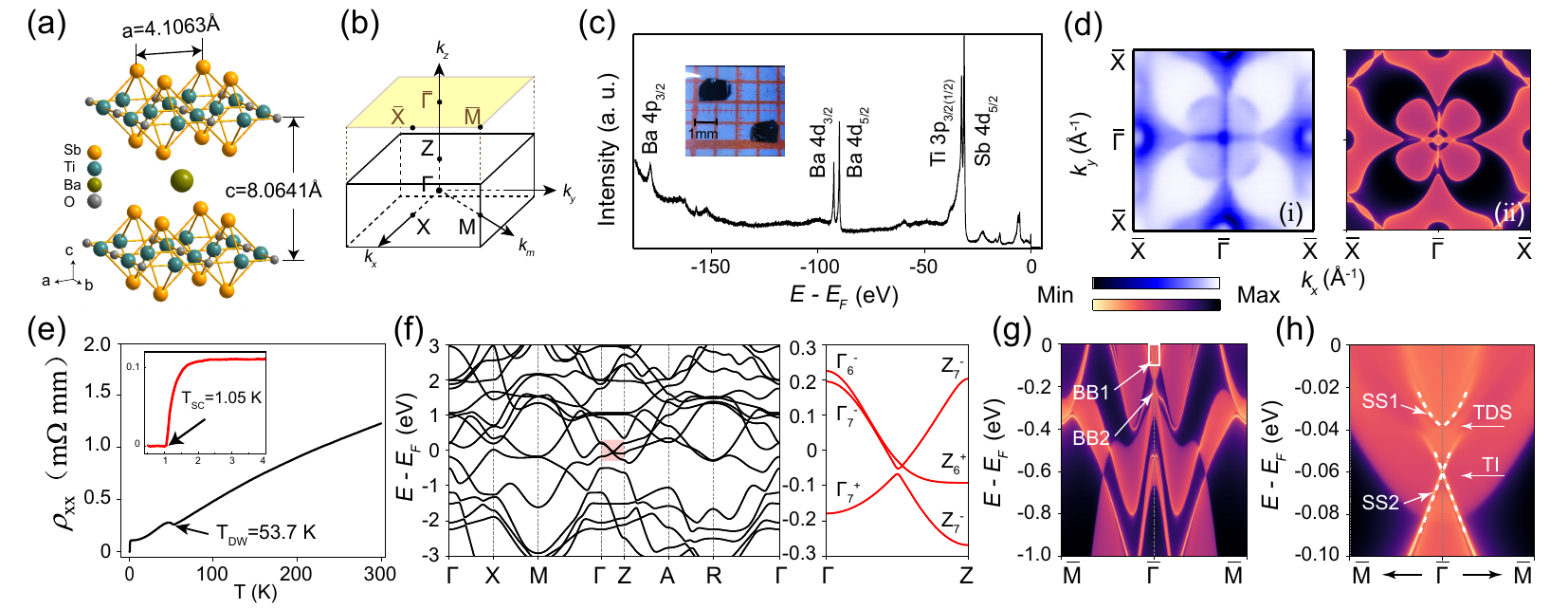}
\caption{(a) Crystal structure of layered BaTi$_2$Sb$_2$O. (b) Three dimensional bulk Brillouin zones (black) with (001) surface Brillouin zone (yellow) projected. High symmetry points are marked. (c) Core-level photoemission spectrum shows characteristic Ba, Ti and Sb peaks. Inset: Image of high quality BaTi$_2$Sb$_2$O single crystal. (d)($\romannumeral1$) and ($\romannumeral2$) Comparison between the experimental and bulk calculated constant-energy surface. ($\romannumeral1$) is taken at 84 eV photons integrated over the energy window of [\emph{E$_F$} - 25 eV, \emph{E$_F$} + 25 eV]. (e) Temperature dependence of resistivity of BaTi$_2$Sb$_2$O. (f) Calculated band structure in the presence of SOC. The enlarged image of the light red rectangle announces the TDS and TI states. (g) Calculated band dispersions along $\bar{\varGamma}$-$\bar{\emph{M}}$ direction. Two bulk bands around $\bar{\varGamma}$ point are marked as BB1 and BB2. (h) Zoom-in of the region in the white box. Two surface states marked as SS1 and SS2 are part of the Dirac cone from TDS and TI, respectively.}
\label{crystal structure}
\end{figure*}

The recent discovery of TI states in the iron-based superconductor Fe(Te,Se), combining nontrivial topological states and superconductivity in a single material, pointed out a new dimension in realizing Majorana bound states~\cite{zhang2018observation,wang2018evidence,gray2019evidence,machida2019zero,kong2019half,wang2020evidence,zhu2020nearly}. One later research on Li(Fe,Co)As revealed that iron-based superconductors might generically host multiple types of nontrivial topological states, e.g., both the TI and topological Dirac semimetal (TDS), together with unconventional superconductivity~\cite{zhang2019multiple}. Thus, the possible intrinsic TSC therein, which takes advantage of the proximity effect in the momentum space, would overcome disadvantages in other implementations of TSC. However, the Dirac point (DP) from the TDS states of superconducting Li(Fe,Co)As with 3\% Co is located above the Fermi level ($E_F$), and thus the proposed one-dimensional Majorana fermion would not be expected to dominate the low-energy electronic structure~\cite{zhang2019multiple}. Meanwhile, seemingly the superconductivity would be inexorably suppressed by introducing further charge carriers in Li(Fe,Co)As~\cite{aswartham2011suppressed,pitcher2010compositional}. In this regard, it is critical to search for more practical superconductors with both Dirac cones from TI and TDS states below $E_F$ to realize multiple topological superconducting states in one material.
 
In this Letter, we have identified both TI and TDS states reminiscent of those in iron-based superconductors but with Dirac cones below $E_F$ in superconducting BaTi$_2$Sb$_2$O samples using angle-resolved photoemission spectroscopy (ARPES) and first-principles calculations. Furthermore, spin-resolved ARPES measurements confirmed both of the predicted spin-helical surface bands originated from TDS and TI states, which are prospective to harbour Majorana zero modes and Majorana flat bands in one single material. This titanium-based oxypnictide superconductor which has the similar multiple topological states as iron-based superconductors would provide another parallel but more practical playground for comparative study on TSC.

\begin{figure*}
\includegraphics[width=17cm]{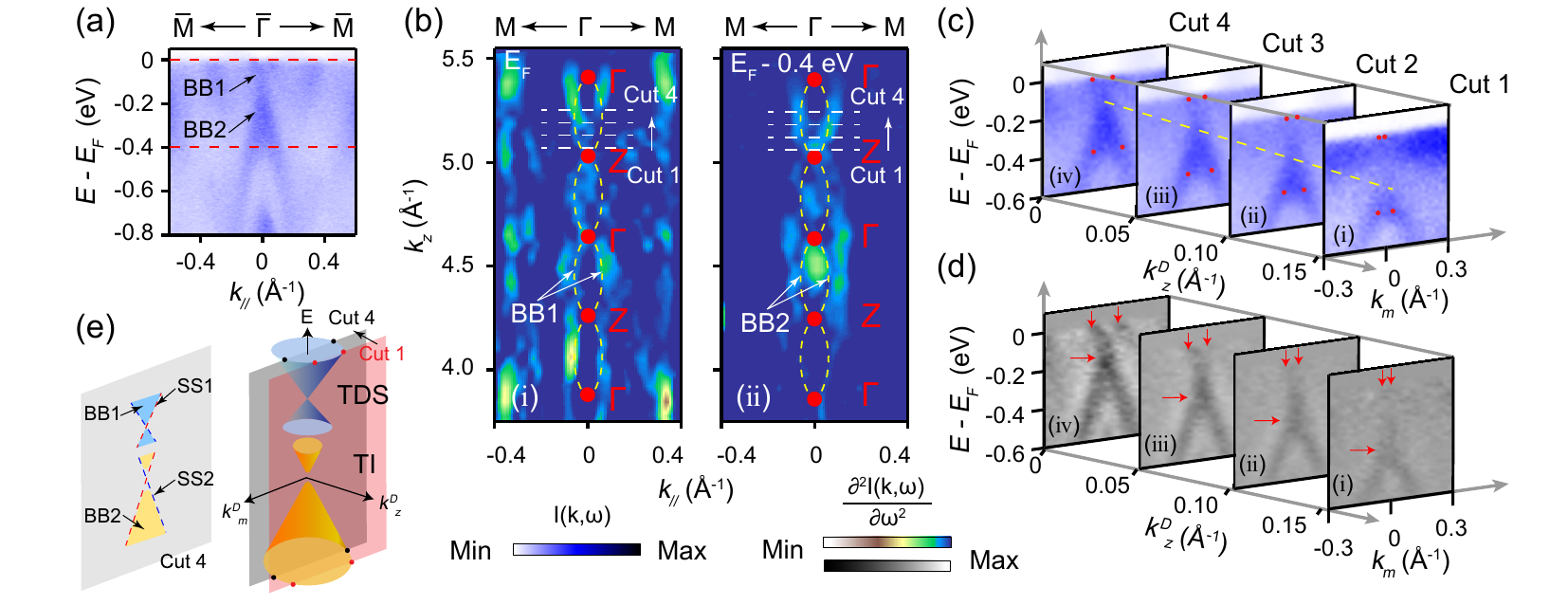}
\caption{(a) Intensity plot along $\bar\varGamma$-$\bar{\emph{M}}$ direction with \emph{p}-polarized 100 eV photons. Two bulk band (marked as BB1 and BB2) near $\bar\varGamma$ point are dominated. (b) Second derivative plot of \emph{$k_z$} dependent ARPES data along $\varGamma$-$\emph{M}$ direction with \emph{p}-polarized photons at \emph{$E_F$} ($\romannumeral1$) and \emph{$E_F$} - 0.4 eV ($\romannumeral2$), which indicate the \emph{$k_z$} dispersion of BB1 and BB2, respectively. The yellow dashed lines are guide to corresponding periodic dispersion along \emph{$k_z$} direction. (c) ARPES intensity plot along $\bar\varGamma$-$\bar{\emph{M}}$ direction at different \emph{$k_z$} indicated by white dashed lines marked as Cut 1 to 4 in (b) with \emph{h}$\nu$ = 94, 96, 98 and 100 eV \emph{p}-polarized photons. (d) Second derivative plot of (c). Two vertical red arrows indicate the evolution of the BB1 and the horizontal arrows indicates the evolution of the BB2. (e) Schematic of projected two Dirac cone from TDS and TI into the (\emph{k$_m$}, \emph{k$_z$}, E) space. The translucent planes indicate the ARPES measurements that intersect through the Dirac cone at different \emph{$k_z$} positions.
}
\label{bulk states}
\end{figure*}


Details on the sample growth method, first-principles calculation and ARPES measurement and can be found in Note 1 to 3 in Supplemental Material (SM)~\cite{si}. The crystal structure of BaTi$_2$Sb$_2$O is illustrated in Fig.~\ref{crystal structure}(a). It is composed of [Ti$_2$Sb$_2$O]$^{2+}$ octahedron layers, which are stacked with Ba atoms along the \emph{c} axis. Thus, the natural cleavage plane should be parallel to the \emph{a-b} plane and between two neighbouring [Ti$_2$Sb$_2$O]$^{2+}$ layers. The bulk and (001)-projected surface Brillouin zones (BZs) of BaTi$_2$Sb$_2$O are shown in Fig.~\ref{crystal structure}(b). Fig.~\ref{crystal structure}(c) displays the angle-integrated photoemission spectrum of BaTi$_2$Sb$_2$O over a large range of binding energy, in which we can clearly identify the Ba (\emph{4p} and \emph{4d}), Ti (\emph{3p}) and Sb (\emph{4d}) core levels, confirming the element composition of our samples. After cleaved in the air, the sample shows typical flat and shining surface as illustrated in the inset of Fig.~\ref{crystal structure}(c). Besides, we present the Fermi surface (FS) map obtained by ARPES in Fig.~\ref{crystal structure}(d)($\romannumeral1$), which clearly suggests the square BZ with four-fold symmetry and is in remarkable agreement with the calculation [Fig.~\ref{crystal structure}(d)($\romannumeral2$)], further proving its tetragonal crystal structure and (001) cleavage plane. In addition, the sample exhibits a metallic temperature dependence and shows zero resistance below $T_c$ = 1.05 K [the inset of Fig.~\ref{crystal structure}(e)], in accordance with previous reports~\cite{yajima2012superconductivity}.

Fig.~\ref{crystal structure}(f) declares the calculated band structure of BaTi$_2$Sb$_2$O with the spin-orbit coupling (SOC). We concentrate on crossings of several bands around $E_F$, which are highlighted by the light red rectangle with the enlarged image in the right panel. These three bands belong to irreducible representations $\Gamma_{6}^{-}$, $\Gamma_{7}^{-}$, and $\Gamma_{7}^{+}$, respectively. The crossing of $\Gamma_{6}^{-}$ and $\Gamma_{7}^{-}$ at an arbitrary $k_{z}$ along $\varGamma$-\emph{Z} features a symmetry-protected DP. Slightly lower in energy, $\Gamma_{7}^{-}$ and $\Gamma_{7}^{+}$ cross, leading to a small gap. The distinct behavior of these two band crossings stems from different mechanisms. BaTi$_{2}$Sb$_{2}$O belongs to the space group no. 123, which respects the inversion symmetry $\hat{\cal I}$. The joint operation of time-reversal ($\cal\hat T$) and $\hat{\cal I}$ promises the Kramer's double degeneracy everywhere in the BZ. In this case, $\hat{c}_{4}$, which leaves $k_{z}$ invariant along $\varGamma$-\emph{Z}, protects a stable DP between $\Gamma_{6}^{-}$ and $\Gamma_{7}^{-}$ bands. However, as $\Gamma_{7}^{-}$ and $\Gamma_{7}^{+}$ share the same basis functions with the only difference on their response to $\hat{\cal I}$, the two bands will unavoidably open a gap when they cross. Such particular alignment of bands is essential for BaTi$_{2}$Sb$_{2}$O to resemble the electronic structure of iron-based superconductors which features a great potential to coexist two distinct types of topological superconductivity in one system, i.e., the one-dimensional Majorana fermion from topological DPs and Majorana zero modes from the topological insulator or DPs~\cite{zhang2019multiple,PhysRevB.100.094520}. To better understand the different symmetry-protection and gap opening mechanisms, we compose a $k\cdot p$ model around the DP and the gap below with $\hat{c}_{4}$ and $\hat{\cal T}\hat{\cal I}$, see Note 4 in SM for more details~\cite{si}.

The above two topological nontrivial states can be further displayed in the surface states calculation along $\bar{\varGamma}$-$\bar{\emph{M}}$. Here, two bulk bands near $\bar{\varGamma}$ are marked as BB1 and BB2, respectively [Fig.~\ref{crystal structure}(g)]. Moreover, by zooming in the region of the white box, we can identify two surface cone-like states, as highlighted by dashed lines in Fig.~\ref{crystal structure}(h). The upper one originates from the TDS and the lower one should be the TI surface state. Although these two DPs are as close as $\sim$23 meV and their overlapped states have gradually merged into bulk states, we could still resolve the rest parts of these surface states, which can be assigned as SS1 and SS2, respectively. We note that these surface states are both below $E_F$ and thus can be probed by ARPES without further carrier doping.

\begin{figure*}[htbp]
\centering
\includegraphics[width=17cm]{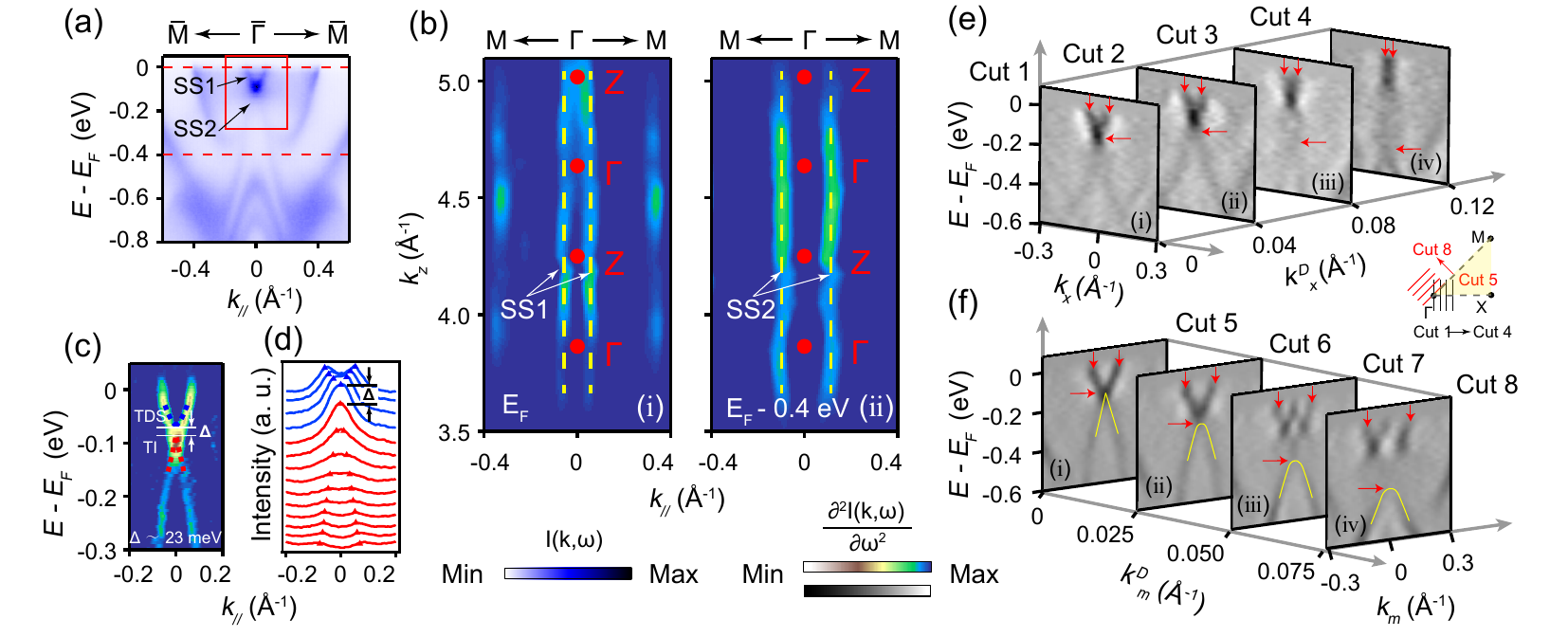}
\caption{(a) Intensity plot along $\bar\varGamma$-$\bar{\emph{M}}$ direction with \emph{s}-polarized 52 eV photons. Two sharp surface states (SS1 and SS2) near $\bar\varGamma$ point are dominated. (b) Second derivative plot of \emph{$k_z$} dependent ARPES data along $\varGamma$-$\emph{M}$ direction with \emph{s}-polarized photons at \emph{$E_F$} ($\romannumeral1$) and \emph{$E_F$} - 0.4 eV ($\romannumeral2$), which indicate the \emph{$k_z$} dispersion of SS1 and SS2, respectively. (c) Second derivative plot of the region in the red box in (a). A gap between two surface Dirac points is marked. (d) MDCs of the region in the red box in (a). The triangles trace the MDC peaks from TDS and TI surface states. (e) Second derivative intensity plots taken at \emph{k$^D_x$} = 0, 0.04, 0.08 and 0.12 \AA$^{-1}$, respectively. (f) Second derivative intensity plots taken at \emph{k$^D_m$} = 0, 0.025, 0.05 and 0.075 \AA$^{-1}$, respectively. Cuts 1 to 8 are illustrated.
}
\label{surface states}
\end{figure*}

Experimentally, we first revealed both the TDS and TI bulk states of BaTi$_{2}$Sb$_{2}$O. Due to considerable entanglement between the bulk and surface states around $E_F$ [Fig.~\ref{crystal structure}(g)], we took advantage of the matrix element effect to distinguish between them by using \emph{p}- and \emph{s}-polarized photons~\cite{zhang2018observation,damascelli2003angle}. The detailed experiment geometry is shown in Fig. S2 and Note 5 in SM~\cite{si}. Fig.~\ref{bulk states}(a) exhibits the photoemisson intensity plot along $\bar{\varGamma}$-$\bar{\emph{M}}$ with \emph{p}-polarized 100 eV photons, corresponding to the $k_z$ plane nearly intersecting both predicted DPs [see details in Fig. S3 and Note 6 in SM] ~\cite{si}. We can directly identify the predicted BB1 (TDS) and BB2 (TI) bulk bands around the BZ center. Figs.~\ref{bulk states}(b)($\romannumeral1$) and ($\romannumeral2$) present second derivative photoemission intensity maps on the \emph{k$_m$-k$_z$} plane [the \emph{k$_m$} direction is indicated in Fig.~\ref{crystal structure}(b)] taken with \emph{p}-polarized photons at \emph{$E_F$} and \emph{$E_F$} - 0.4 eV, respectively. Both band dispersions with apodictic periodic modulation along \emph{$k_z$} could be recognized, confirming the bulk nature of BB1 and BB2. Furthermore, we show four photoemission intensity plots taken at different $k_z$ as schematically illustrated in Fig.~\ref{bulk states}(e), from Cut 1 [Fig.~\ref{bulk states}(c)($\romannumeral1$)] to Cut 4 [Fig.~\ref{bulk states}(c)($\romannumeral4$)]. The dispersion of BB2 varies from the parabolic lineshape to quasi-linear, and the apex keeps drifting up (highlighted by the yellow dashed line), showing the typical feature of a TI bulk cone [Fig.~\ref{bulk states}(e)]. In addition, the corresponding second derivative plots in Fig.~\ref{bulk states}(d)($\romannumeral1$)-($\romannumeral4$) manifest the gradual increase of the Fermi crossing and more linear dispersion when close to the DP, in accordance with the predicted TDS feature of BB1 [Fig.~\ref{bulk states}(e)].

Next, we performed similar ARPES measurements but with \emph{s}-polarized photons to identify topological surface states of BaTi$_{2}$Sb$_{2}$O. Fig.~\ref{surface states}(a) shows the intensity plot along $\bar{\varGamma}$-$\bar{\emph{M}}$ with \emph{s}-polarized photons. The predicted surface bands SS1 and SS2 around $\bar{\varGamma}$ were unambiguously revealed. Figs.~\ref{surface states}(b)($\romannumeral1$) and ($\romannumeral2$) represent second derivative intensity maps on the \emph{k$_m$-k$_z$} plane with \emph{s}-polarized photons taken at \emph{$E_F$} and \emph{$E_F$} - 0.4 eV, respectively. Our results demonstrate negligible \emph{k$_z$} dispersions for both bands, suggesting that bulk states have been significantly suppressed and surface states from TDS and TI are dominating 
here. Fig.~\ref{surface states}(c) displays the enlarged second derivative plot of band structure in the red box of Fig.~\ref{surface states}(a). Together with the appended band dispersions (blue and red dashed lines) extracted from the corresponding momentum distribution curves (MDCs) [Fig.~\ref{surface states}(d)], we can well distinguish between those cone-like surface states from TDS and TI, in spite of rather small energy gap between these dual topological states. Furthermore, as exhibited by second derivative intensity plots in Figs.~\ref{surface states}(e) and (f), detailed evolution of these surface cones along \emph{k$_x$} and \emph{k$_m$} could be recognized in evidence. The in-plane dispersion migrates apparently deviating $\varGamma$ at different \emph{k$_x$} and \emph{k$_m$} positions, suggesting that both surface DPs are exactly located at $\bar\varGamma$.

\begin{figure*}[htbp]
\centering
\includegraphics[width=17cm]{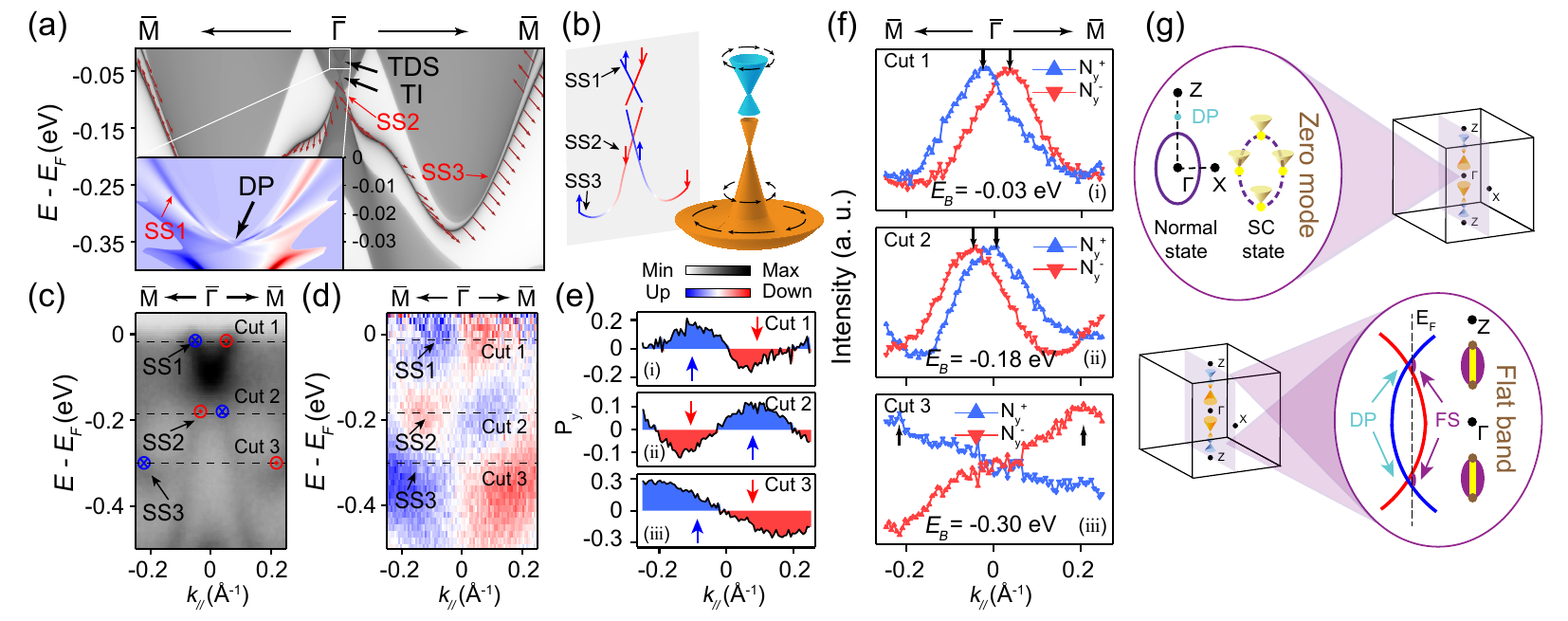}
\caption{(a) Calculated spin texture along $\bar{\varGamma}$-$\bar{\emph{M}}$ direction. Inset: the enlarged image of the white box region. (b) Sketch of the spin-polarized characteristics of TDS and TI surface states. The red and blue lines express spin down and up, respectively. (c) and (d) Intensity plot measured without and with $\hat{y}$ direction spin-polarized along $\bar{\varGamma}$-$\bar{\emph{M}}$ direction. The red and blue parts express spin direction along -$\hat{y}$ and +$\hat{y}$, respectively. (e) Spin polarization at Cut 1 to 3 in (d), indicating the spin texture of SS1 to SS3. (f) Spin resolved MDCs taken at Cut 1 to 3. The red (down) and the blue (up) triangles indicate the spin direction along -$\hat{y}$ and +$\hat{y}$,respectively. (g) Sketch of the dual topological superconductivity states induced by TDS states.}
\label{spin texture}
\end{figure*}

Both surface cones originating from TI and TDS should be spin-polarized, as suggested by our first-principles calculations [Fig.~\ref{spin texture}(a) and the inset]. Surface states from the upper part of TDS cone (SS1), the lower part (SS2) and extension of TI cone (SS3) all suggest the same characteristic left-hand spin helicity as Li(Fe,Co)As and most TIs, which is then further confirmed by our constant-energy spin texture calculations shown in Fig. S4 and Note7 in SM~\cite{si}. We then used spin-resolved ARPES to check these predictions. Here, for easy comparison, we summarized the sketch of spin nature of SS1 to SS3 in Fig.~\ref{spin texture}(b) according to calculations. Figs.~\ref{spin texture}(c) and (d) show our photoemission intensity plots without and with $\hat{y}$ direction spin-resolved measurement in the same region under more surface sensitive \emph{s}-polarized photons, respectively. Evidently, for both TI and TDS surface cones, the left hand parts of these cones show the opposite spin direction to those of right hand parts, consistent with our calculations. Moreover, as displayed in Figs.~\ref{spin texture}(e), spin-resolved MDCs taken at \emph{E$_B$}  = -0.03 eV (Cut 1), -0.18 eV (Cut 2) and -0.30 eV (Cut 3) intersecting SS1 to SS3 demonstrate strong spin polarization up to 0.3. Meanwhile, MDCs of partial intensities along -$\hat{y}$ and +$\hat{y}$ at Cut 1 to 3 in Fig.~\ref{spin texture}(f) also support the spin polarization result [see more details in Fig. S5 and Note 8 in SI]~\cite{si}. We note that these MDCs' peaks (marked as black arrows) indicate the exact momentum of surface bands, which are consistent with general ARPES results in Fig.~\ref{spin texture}(c).

The coexistence of spin-helical topological surface states and the bulk TDS in BaTi$_{2}$Sb$_{2}$O resembles that of Li(Fe,Co)As. The DPs formed by the band inversion between Ti-$d$ and Sb-$p$ stay slightly below $E_F$, which naturally forms a FS with parity mixing - a necessary ingredient for the emergent TSC in TDS~\cite{PhysRevB.100.094520}. On the side surface, such as $(010)$ schematically shown in Fig.~\ref{spin texture}(h), the bulk DPs map to different surface nodes and there may exist Fermi arcs/loops between the two DPs~\cite{yang2014classification}. They would be gapped by the proximity to the superconductivity except for the 
four nodes along $\varGamma$-\emph{Z} and $\varGamma$-\emph{X}, forming a zero energy Majorana mode. More interestingly, the parity-mixing Fermi surface above the DPs, as shown by the violet sphere in Fig.~\ref{spin texture}(h), would be gapped as well. However, as long as the surface admits the mirror symmetry along $y$-direction, Majorana flat-bands will appear as symmetry-protected TSC. Thus, BaTi$_{2}$Sb$_{2}$O, not belonging to any existing iron-based superconductor family, provides another exciting platform for exploring the proximity effect in momentum space, which will greatly enrich the topological TSC family and offer independent insights for examining the current understanding of the TSC. Furthermore, compared to Fe(Te,Se) and Li(Fe,Co)As, as no carrier doping is required in BaTi$_2$Sb$_2$O, the perfect alignment of $E_F$ and DPs makes it the so-far most promising candidate to detect multiple-type TSC in one system, including the Majorana zero mode and the surface flat-bands. 

In fact, $E_F$ of BaTi$_2$Sb$_2$O can be easily tuned through the Na content, and the TI/TDS bands can be separately accessed. Since both these TI/TDS bands are dominated by Ti orbitals, the doped Na would change the dispersion negligibly. Our Na doping results [Fig. S6 and note 9 in SM] and other previous works have given an explicit adjustment of $E_F$ without alteration of TI/TDS non-trivial topological states around $\varGamma$ point~\cite{song2016electronic,si}. Furthermore, the superconducting transition temperature can be tuned consequently by Na doping, and the Ba$_{1-x}$Na$_x$Ti$_2$Sb$_2$O (0 $\le$ x $\le$ 0.33) is expected to achieve a maximum superconductivity critical temperature $T_c$ = 5.5 K \cite{doan2012ba1}.



In summary, we have identified the symmetry protected TDS, the TI states and their spin-momentum locking patterns in the titanium-based oxypnictide superconductor BaTi$_2$Sb$_2$O by ARPES measurements in association with first-principles calculations. The helical surface states from TDS which is in the vicinity of \emph{E$_F$} could produce the Majorana zero mode and Majorana flat band in this system. Therefore, BaTi$_2$Sb$_2$O is a good platform to obtain various types of Majorana fermions and further research of topological superconductivity.

\begin{acknowledgments}
This work was supported by the National Key R$\&$D Program of the MOST of China (Grants No. 2016YFA0300204, 2017YFE0131300 and 2016YFA0401002), the National Science Foundation of China (Grant Nos. 11404360, 11874264, 11574337 and 11704394), the Strategic Priority Research Program of Chinese Academy of Sciences (Grant No. XDA18010000) and the Natural Science Foundation of Shanghai (Grant No. 14ZR1447600). Y.F.G. and G.L. acknowledge the starting grant of ShanghaiTech University and the Program for Professor of Special Appointment (Shanghai Eastern Scholar). Y.B.H. acknowledges the CAS Pioneer ``Hundred Talents Program" (type C). D.W.S. is supported by ``Award for Outstanding Member in Youth Innovation Promotion Association CAS''. Part of this research used Beamline 03U of the Shanghai Synchrotron Radiation Facility, which is supported by ME2 project under contract No. 11227902 from National Natural Science Foundation of China. Calculations were carried out at the HPC Platform of ShanghaiTech University Library and Information Services, and at School of Physical Science and Technology. The authors also thank the support from the Analytical Instrumentation Center (SPST-AIC10112914), SPST, ShanghaiTech University. 

\end{acknowledgments}


\bibliographystyle{apsrev4-1}

\end{document}